\documentclass[reqno]{amsart}
\usepackage{graphics,amsmath}

\newcommand{\xt}{x^i}
\newcommand{\pt}{p^i} 
\newcommand{\dxt}{\dot x^i}
\newcommand{\dpt}{\dot p^i} 

\newcount\n  \newdimen\w

\def\Repeat#1#2{\n=#1\relax\loop\ifnum       
  \n>0\relax #2\advance\n by-1\repeat}

\long\def\OMIT#1{\relax }  

\def\re#1{(\ref{#1})}   
  
\def\eqn#1#2{ \begin{align} \label{#1}         #2 \end{align}}

\def\nl#1{          \\ \label{#1}        }  
\def\nnl#1{ \tag*{} \\ \label{#1}        }  

\def\delim#1#2#3{\csname\ifcase#1 relax\or   
   big\or Big\or bigg\or Bigg\fi\endcsname   
  {\ifcase#2\or\Delim#3\or\deliM#3\fi}}      
\def\Delim#1{\ifcase#1\relax\or(\or[\or\{\or<\or\langle\or|\or\|\or---{ }\fi}
\def\deliM#1{\ifcase#1\relax\or)\or]\or\}\or>\or\rangle\or|\or\|\or{ }---\fi}

\def\largerfrac#1#2#3{      
  \whichtypesize\n=\currenttypesize\advance\n by #1 \mathchoice
  {\setbox0\hbox{$\displaystyle-$} \w=.5\ht0\advance\w by-.5\dp0\setbox0
    \hbox{\typesize\n $\displaystyle-$} \advance\w by -.5\ht0\advance\w
    by .5\dp0\raise\w \hbox{\typesize\n$\displaystyle{\frac{#2}{#3}}$}}
  {\setbox0\hbox{$-$} \w=.5\ht0 \advance\w by -.5\dp0 \setbox0\hbox
    {\typesize\n $-$} \advance\w by-.5\ht0\advance\w by
    .5\dp0\raise\w\hbox{\typesize\n$\frac{#2}{#3}$}}
  {\setbox0\hbox{$\scriptstyle-$} \w=.5\ht0 \advance\w by-.5\dp0\setbox0
    \hbox{\typesize\n $\scriptstyle-$} \advance\w by -.5\ht0 \advance\w
    by .5\dp0 \raise\w\hbox{\typesize\n$\scriptstyle{\frac{#2}{#3}}$}}
  {\setbox0\hbox{$\scriptscriptstyle-$} \w=.5\ht0
    \advance\w by -.5\dp0 \setbox0\hbox{\typesize\n
    $\scriptscriptstyle-$} \advance\w by -.5\ht0 \advance\w by .5\dp0
    \raise\w\hbox{\typesize\n$\scriptscriptstyle{\frac{#2}{#3}}$}}  }

\def\d{{\rm d}}       

\def\Laplace{\bigtriangleup}  
\begin{document}

\title{Variational principles and thermodynamics}
\author{P. V\'an$^{1,2,3}$  and R. Kov\'acs$^{1,2,3}$  }
\address{$^1$Department of Theoretical Physics, Wigner Research Centre for Physics, H-1525 Budapest, Konkoly Thege Miklós u. 29-33., Hungary; //
and  $^2$Department of Energy Engineering, Faculty of Mechanical Engineering,  Budapest University of Technology and Economics, 1111 Budapest, Műegyetem rkp. 3., Hungary//
$^3$ Montavid Thermodynamic Research Group}
 
\date{\today}

\begin{abstract}
Variational principles play a fundamental role in deriving evolution equations of physics. They are working well in case of nondissipative evolution but for dissipative systems they are not unique, not predictive and not constructive. With methods of modern nonequilibrium thermodynamics, one can derive evolution equations for dissipative phenomena and, surprisingly, can also reproduce the Euler-Lagrange form of the evolution equations for ideal processes. In this work, we examine some demonstrative examples and compare thermodynamic and variational techniques. Then, we argue that instead of searching for variational principles for dissipative systems, a different program can be more fruitful: the second law alone can be an effective tool to construct both dissipative and nondissipative evolution equations.
\end{abstract}
\maketitle


\section{Variational principles of dissipative systems}
Variational principles are the ultimate tools for deriving the evolution equations of physics, in particular the equations of motions in mechanics. They work well for ideal systems, when the evolution is based on symmetric differential operators \cite{Vai64b,MorFes53b}. Variational principles for dissipative systems are special, modifications are necessary to circumvent the strict mathematical conditions \cite{YouMan99b,FinScr67a,SieFar05b}. {The applied tricks can be rather different, for instance  changing the original equations, increasing the number of variables, reducing the corresponding function spaces, turning to statistical interpretations and so on \cite{Gya70b,Vuj89b,EngYah04a,Pol14a1,Gla15a,Abe19mtv}}. 

These methods are special, different and not equivalent \cite{VanNyi99a}. A general construction methodology of dissipative evolution equations in physics is still a challenge. Any candidate principle, variational or not, should be able to handle both dissipative and nondissipative evolution equations. For the ideal case it must be compatible with Hamiltonian principles, and in the nonideal case, with the principles of thermodynamics. Moreover, it should be universal, and, at the same time, using only a minimal number of assumptions. 

In this paper, we reverse the usual logic. Instead of extending the reach of variational principles to dissipative systems, we argue that the second law can be used to derive evolution equations for both dissipative and nondissipative evolution. First, recent developments of nonequilibrium thermodynamics are outlined. Then the evolution equation of a single point mass is treated. Here, one need dual thermodynamic variables. After that, a more complicated example, a two-component compressible fluid is considered, where constraints and weak nonlocality lead to the Fourier--Navier-Stokes--Cahn-Hilliard--Korteweg equations. There a weakly nonlocal extension of the entropy principle is used. In the fourth section, with a combination of dual variables and weak nonlocality,  an inertial extension of the classical gravitation for self-gravitating heat conducting fluids is derived. Finally, we summarize, compare and discuss the results. 

\section{Gradients and internal variables in nonequilibrium \\ thermodynamics}

In classical nonrelativistic irreversible thermodynamics one calculates the entropy balance in order to obtain constitutive functions that are compatible with the second law \cite{GroMaz62b}. There the basic assumption is called the hypothesis of local equilibrium.  In modern theories of nonequilibrium thermodynamics, both weak memory and weak nonlocal effects are treated, through internal variables and gradients, respectively. Both aspects are reflected in methods exploiting the entropy inequality, and must consider frame indifference and objectivity as a fundamental requirement \cite{Fre09a,Mar16a}. 

The second law, the entropy inequality, is conditional. An absolute, i.e, frame independent treatment requires a four dimensional spacetime representation of the physical quantities also in a nonrelativistic case. In this work we consider the following aspects of a more rigorous nonrelativistic formalism \cite{Mat93b,Van17a}:
\begin{itemize}
\item Both  {the} entropy density and  {the} entropy flux are constitutive quantities,  {they are} the  {time-like} and  {space-like} components of the entropy four vector. Therefore {, the} entropy flux is to be determined according to the second law. 
\item Gradients are invariant, frame independent quantities, because a gradient is a nonrelativistic spacelike covector. Therefore gradient dependent constitutive functions are frame independent. Time derivatives are not objective. 
\end{itemize}

Using these simple rules, objectivity and frame independence can be incorporated in the theory, without any formal, transformation rule based calculations \cite{MatVan06a,Van17a}. Therefore in this short communication we can keep a familiar, simple presentation form, while keeping the objectivity and frame indifference of the continuum theory. Also methods which are rigorously exploiting the entropy inequality, called Coleman-Noll and Liu procedures, can be extended to deal with weak nonlocality in time \cite{VanAta08a,BerVan17b}  and in space \cite{Van04a,Cim07a}, while respecting frame independence.

In Extended Thermodynamics, the dissipative fluxes are not constitutive functions. They are incorporated among the basic state variables as new fields \cite{JouAta92b,MulRug98b}. Then the evolution equations of these new fields are to be constructed according to the requirements of the second law \cite{Gya77a}.  {In general}, the idea   {to extending the} state with new fields, that of internal variables, is not new \cite{Duh11b,ManLeo37a,Bri43b,Mau99b}. There are many interesting applications in their thermodynamically consistent use with  {particular physical examples} \cite{MauMus94a1,MauMus94a2,Ver97b}. Their combination with the classical fields understood as memory effects due to internal structural changes:  {their elimination results in} higher order time derivatives  {in} the original evolution equations, they represent an inertial memory effect, that is weak nonlocality in time.

There are many different methods to introduce weak nonlocality in space, that is, extending the constitutive functions with the gradients of the thermodynamic state variables, both   {for} classical   {and} internal ones. Without  completeness, one need to mention the method of virtual power \cite{Mau80a,Var11b,For19mtv}, GENERIC \cite{GrmOtt97a,Ott05b,PavEta18b}, configurational forces \cite{Gur00b,Mau11b}, conservation-dissipation formalism \cite{Yon19tvm} and phase field theories \cite{HohHal77a}. All of these approaches can produce successful theories, with various levels of thermodynamic consistency and objectivity. For example, phase field theories introduce functional derivatives for the new fields and then the variational techniques are combined with thermodynamic methods. 

There are some particular aspects for the correct application of weak time and space nonlocality in a thermodynamic framework.
\begin{itemize}
\item Internal variables are incorporated in the thermodynamic state space, i.e., the entropy density may depend on them. The related Gibbs relation is a convenient tool in calculations.
\item The entropy density may depend on the gradients of the state variables, including internal ones (contrary to \cite{ColGur67a}), too.
\item A field theoretical approach is preferable, that is, the densities or specific quantities are primary in a continuum theory. That  {holds} for the entropy, too.  {Consequently,} the first order Euler homogeneity of the entropy applies for internal variables and gradient extensions. 
\end{itemize}

In the following, we demonstrate that the exploitation of the second law for internal variables with weakly nonlocal, gradient dependent state spaces is a constructive method to obtain the evolution equations of a theory, both  {for dissipative and nondissipative situations.} We  {are presenting three} examples. First, the appearance of symplectic structures, with dual internal variables is demonstrated in point mass mechanics. Then, the use of gradient methods is shown on the example of two-component compressible and diffusive Fourier--Navier-Stokes--Cahn-Hilliard--Korteweg fluids, which is a complicated, but typical example when the phase field methodology with variational derivative problematic. Our final example is the derivation of equations of classical gravitation with inertia, applying the same treatment. It is a combination of internal variables and classical fields both with gradient and memory effects.

In all these cases, our concepts are  based on the universality of a thermodynamic approach. We do not   {intend to} look for the microscopic or mesoscopic origin of the fields. They may be considered as emergent, but here it is not important, the thermodynamic treatment is uniformly   {applicable, even} when the usual approach is not statistical, e.g. for point masses or gravitation. 

{From now on}, we use   {the abstract index notation} with $i,j,k$   {being spatial indices}, denoting the tensorial character of a physical quantity in a three dimensional vector space without particular coordinates \cite{Pen04b}. Identical upper and lower indices denote contraction, according to the summation rule.

\section{Example I: Thermodynamic approach to mass point mechanics}

{The} Newtonian, Lagrangian and Hamiltonian mechanics of a single point mass, \cite{Arn78b}, {are solely} obtained from thermodynamic considerations, and compared to the evolution equations obtained from   {mechanics}. This particular thermodynamic point of view leads to   {the} dissipative version of Hamiltonian mechanics .

\subsection{Pure mechanics}
The Hamilton equations of a point mass are generated from a potential called Hamiltonian, $H(\xt,\pt)$, and are given as
\eqn{h1}{
\dxt = \partial_{\pt} H, \nl{h2}
\dpt = -\partial_{\xt} H
}
  {with the upper dot denoting the time derivative}.

A partial Legendre transformation gives the Lagrangian $L(\xt,\dxt)$ as:
\eqn{lmtraf}{
L(\xt,\dxt) + H(\xt,\pt) = \pt \dot x_i.
}
The derivatives of this formula   {yield}
\eqn{lc1}{
\partial_{\dxt} L = \pt, \nl{lc2} 
\partial_{\pt} H = \dxt, \nl{lc3} 
\partial_{\xt} L = -\partial_{\xt} H.
}
The equation of motion, the Euler-Lagrange equation follows by eliminating $\pt$, the momentum from the system in order to obtain a differential equation for the position $\xt$:
\eqn{ELpmech}{
\frac{d}{dt}\partial_{\dxt}L -\partial_{\xt} L= \dpt + \partial_{\xt} H = 0.
}

This formula is obtained with the usual interpretation that \re{h1} (and \re{lc2}) is the definition of the momentum, and \re{h2} is the equation of motion. The use of a Lagrangian is a convenient way to perform the elimination through the Legendre transformation \re{lmtraf}, without directly inverting the Hamiltonian. 

{Eq.~\re{ELpmech} is an evolution equation without dissipative terms.} Those are of secondary importance in the usual approach and given by separate, additional assumptions, such as imitating the nondissipative mathematical structure by dissipation potentials of Rayleigh.  {That is, the addition of dissipative terms requires additional concepts}, e.g. dissipation potentials. Without introducing thermodynamics one cannot be sure whether a modification of the ideal equation is correct or not.  {For instance,} in an additive  damping term like $\beta\dot{\xt}$, the coefficient $\beta$ must have a definite sign.

\subsection{Pure thermodynamics}

The thermodynamic construction of an evolution equation for  {internal state variables is} also simple. One assumes the existence of a concave potential, which is increasing  {in time}. Let us observe the direction of argumentation: one does not start from the evolution equations, but from the potential  {that} determines the evolution according to the thermodynamic requirements. It is important to note: \emph{these two conditions define a thermodynamic system}  {hereafter}.  {Let us denote the state variable by} $\pt$, and the potential by $S$, then 
\eqn{srel}{
\dot S(\pt) = \partial_{\pt} S \dpt \geq 0,
}
therefore, in the simplest case,  {when} $S = -\frac{p^2}{2m}$ with $m>0$, then a linear isotropic solution of the above inequality leads to 
\eqn{releq}{
\dpt = - \frac{l}{m} \pt, 
} 
where $l>0$. It is a relaxation dynamics, the concavity of the potential ($m>0$) together with  {the} condition \re{srel}  {guarantee} the asymptotic stability of the equilibrium, $\pt =0$.  This is what one expects from a thermodynamic evolution: the hot coffee cools to the temperature of the environment, the asymptotic stability of the equilibrium is ensured.  {Indeed}, the time dependent classical thermodynamics of homogeneous systems rigorously and generally satisfies that expectation as it is proven by Matolcsi \cite{Mat05b}. One may think that thermodynamics excludes inertial effects and for mechanics, or for analogous evolutions with inertia, a variational principle is necessary, as it was argued e.g. in \cite{Mau90a,MauMus94a1}. However, it is not so. 

\subsection{Dual variable thermodynamics}

Inertial effects can be incorporated because dual variables naturally lead to Hamiltonian dynamics if the potential is conserved. Let the state space of a thermodynamic system be spanned by two variables, $\xt$ and $\pt$, and let us consider a concave potential $S(\xt,\pt)$, increasing along the evolution of these state variables. We do not assume any more particular constraints for these physical quantities, therefore, they are internal in the thermodynamic sense.   {Hence}
\eqn{sdyn}{
\dot S(\xt,\pt) = \partial_{\xt} S \dxt + \partial_{\pt} S \dpt \geq 0.
}
If $S$ is a two times differentiable function with continuous second derivative on a simple connected domain, then a quasilinear solution of this inequality  can be written, in general, as:
\eqn{dmech1}{
\dxt &= l_{11} \partial_{\xt} S + l_{12} \partial_{\pt} S,\nl{dmech2}
\dpt &= l_{21} \partial_{\xt} S + l_{22} \partial_{\pt} S.
}
Regarding more precise arguments and conditions see e.g. the Appendix B in \cite{Gur96a}.
Here, the conductivity tensor
\eqn{ctens}{
L = \begin{pmatrix}
l_{11} & l_{12}\\
l_{21} &l_{22}
\end{pmatrix}} 
is not necessarily symmetric, but its symmetric part is positive definite, according to the inequality \re{sdyn}, therefore
\eqn{pdefc}{
l_{11} >0, \quad l_{22} >0, \quad l_{11}l_{22}- \frac{(l_{11}+l_{11})^2}{2}>0.
}
It is convenient to split $L$ into symmetric, $L_D$, and antisymmetric $L_I$, parts as
\eqn{csptens}{
L = L_I + L_D=\begin{pmatrix}
0 & -k\\
k & 0
\end{pmatrix} +
\begin{pmatrix}
l_{11} & l\\
l &l_{22}
\end{pmatrix},
}
where the notation $l = (l_{12}+l_{21})/2$ and $k = (l_{21}-l_{12})/2$ are introduced. Then the third condition in \re{pdefc} is $l_{11}l_{22}- l^2>0$. If the dissipative   {(symmetric)} part is zero,   {the conservation of $S$ follows}. The   {antisymmetric} part $L_I$ generates ideal evolution, a symplectic dynamics. It is identical with \re{h1}-\re{h2} if $k=1$. 

This argument is   {the} cornerstone of metriplectic dynamics \cite{Mor86a,Mat16a} and in single generator GENERIC. In both cases, the dynamics is generated by a gradient of a potential. It is also thermodynamics, $S$ may be like an entropy and the relation to the Hamiltonian $H$ can be  $S(E-H) = - c \ln(E-H)$,   {with} $E$ being the total energy of the point mass and $c>0$ its heat capacity. However, in case of pure mechanics without temperature, $H$ may play the role of free energy and $S:=-H$. This changes the concavity of the potential but the dynamics   {will} be the same. The usual dissipative contributions (damping, friction) can be incorporated in the term $l_{22} \partial_{\pt}S$. However, the other terms also represent various dissipative effects, the question is why we do not realize them? 

It was argued several times that the symplectic structure is the core of mechanics. A Lagrangian can be calculated from a Hamiltonian, and then the classical variational principles can be constructed. However, the Lagrangian and Hamiltonian approaches are not equivalent neither for ideal dynamics and the situation is worse for dissipative systems.

\subsection{Mechanothermodynamics}

Normally, in a textbook treatment, mechanics starts from the Newton equation with nondissipative dynamics. However, in a dissipative system this start can be misleading. As we have already mentioned, the equations \re{dmech1}-\re{dmech2} generalize usual mechanical dissipation. In order to get an impression about the physics of this generalisation, it is worth to consider a specific example, a usual Hamiltonian with kinetic and potential energies, i.e., $S(\xt,\pt) = -H(\xt,\pt) = -\frac{p^2}{2m} - V(\xt)$, where $V$ is a convex function. In this case, \re{dmech1}-\re{dmech2} can be written as
\eqn{sdmech1}{
\dxt &= \frac{k}{m}\pt - l_{11} \partial_{\xt} V - \frac{l}{m}\pt, \nl{sdmech2}
\dpt &= -k \partial_{\xt} V - l \partial_i V - \frac{l_{22}}{m} \pt.
}
Then $\pt$ can be expressed explicitly, and one obtains
\eqn{momm}{
\pt = m\frac{\dxt +l_{11} \partial_{\xt} V}{k-l}.
}
This is a strange, potential dependent `momentum'. Substituting this expression into \re{sdmech2}, one derives the corresponding dissipative equation of motion:
\eqn{seqm}{
\hat m \ddot x^i + \hat m(l_{11} \partial_{x^ix^j}V + l_{22}/m\delta_{ij})\dot x^j + \partial_{\xt}V = 0,
} 
{in which $\hat m = m/(k^2+ l_{11} l_{22}-l^2 )$, ($\hat m >m>0$), all coefficients are nonnegative according to the second law and the convexity of the Hamiltonian}.  We can see that following the usual reasoning, i.e., starting from the equation of motion, the separation of the dissipative and nondissipative parts is not straightforward. The mass increases due to dissipation and the damping is affected by the potential. However, when an ideal, nondissipative motion is what we keep  in our mind, then the simple damping term in the middle of \re{seqm} seems to be completely satisfactory. Also, in ideal motion $k \neq 1$ is invisible, because the momentum is defined from the Newton equation. The interpretation of dissipative terms in \re{h1}, the `phase dissipation' is a peculiar phenomenon. The physical significance is best understood from a stochastic point of view,  for example, it may appear when the position is fluctuating like in case of wave function collapse \cite{Dio88a,Dio15a}.

We have seen that inertial effects can be introduced by pure thermodynamics, and this generalization leads to a fundamental reinterpretation of concepts and expectations for the simplest mechanical system a point mass in an external potential. But aside from the thermodynamics of a point mass, one thing is clear. In order to get Hamiltonian dynamics one does not need a variational principle, but dual variables and conserved entropy.


\section{ {Example II -- Phase fields: Fourier--Navier-Stokes--Cahn-Hilliard--Korteweg equations} }

Classical phase-field models, the Allen-Cahn (or Ginzburg-Landau, model A) and the Cahn-Hilliard (model B) equations were not obtained from a thermodynamic treatment \cite{HohHal77a,Gur96a}.  {As we have already mentioned, there are several different methods to derive them}, with or without variational derivatives \cite{Gur96a,FabEta06a,Paw06a,Gio09a,FreKot15a}. The Cahn-Hilliard equation is exceptionally challenging for rigorous derivation, because of the higher space derivatives \cite{Van18bc}. In this section, we investigate a two-component  compressible heat conducting fluid with diffusion.  {Using} variational principles, it is not easy to realize the balances as thermodynamic constraints, because of the higher  {order} derivatives in the state space, and the nontrivial contributions to the pressure \cite{LowTru98a}. Here we give a simple, but complete treatment,  utilizing the Gibbs relation as a starting point.  {In} the identification of the entropy flux we use the method of separation of divergences \cite{GroMaz62b,Mau06a,Van19m,Van18bc}.

\subsection{Basic balances}

A two-component fluid with component densities $\rho_1, \rho_2$ and velocities $v^i_1, v^i_2$ will be characterized by the  {density $\rho$, the  baricentric velocity $v^i$, the concentration $c$, and the diffusion flux $j^i$. They} defined as 
\eqn{vardef}{
\rho = \rho_1+\rho_2, \qquad
v^i = \frac{\rho_1 v^i_1 + \rho_2v^i_2}{\rho_1+\rho_2}, \qquad
c = \frac{\rho_1}{\rho}, \qquad
j^i = \rho_1 (v^i_1 -v^i).
}
 {The balances of mass and concentration are}
\eqn{mbal}{
\dot\rho + \rho\partial_i v^i &= 0, \nl{cbal} 
\rho\dot c + \partial_i j^i &= 0,
}
 {where the dot denotes the substantial time derivative  and $\partial_i$ is the nabla operator}. The balance of momentum, the Cauchy equation, and the balance of internal energy are
\eqn{mobal}{
\rho \dot v^i + \partial_j P^{ij} &= 0, \nl{ebal}
\rho \dot e + \partial_i q^i &= -P^{ij}\partial_i v_j,
}
where $P^{ij}$ is the pressure tensor and $q^i$ is the heat flux. 

The basic fields of the system are the internal energy, the density, the velocity and the concentration. In order to get a closed system of equations, one  {must} derive the constitutive functions for the diffusion flux $j^i$, the pressure $P^{ij}$, and the heat flux $q^i$.  {These relations can be determined with the help of the entropy inequality}. 

Our thermodynamic state variables are the internal energy, the density and the concentration. The specific entropy is the function of these fields and also depends on the concentration gradient,  {i.e., $s(e,\rho,c,\partial_i c)$, which is required to obtain the necessary couplings.} The Gibbs relation introduces the intensive thermodynamic quantities:
\eqn{CHGibbs}{
de= T\d s - p dv + \mu \d c +\frac{Z^i}{\rho} \d\partial_i c.
}
That is, the partial derivatives of the specific entropy are
\eqn{pards}{
\partial_es = \frac{1}{T}, \quad
\partial_\rho s = -\frac{p}{\rho^2 T}, \quad
\partial_cs = -\frac{\mu}{T}, \quad
\partial_{\partial_ic}s = -\frac{Z^i}{\rho T},
}
where $T$ is the temperature, $v = 1/\rho$ is the specific volume, $p$ is the thermostatic pressure, $\mu$ stands for the chemical potential and  $Z^i$ represents the derivative of the entropy by the concentration gradient. $Z^i$ is introduced respecting the extensivity of the entropy. Now the entropy production can be calculated from the time derivative of the entropy  {considering} the balances \re{mbal}, \re{cbal} and \re{ebal} as constraints, then separating the surface and volumetric effects. This is a short, simple and straightforward  calculation:
\eqn{cCH1}{
\rho \dot s(e,\rho,c,\partial_i c) &= \frac{1}{T}\rho \dot e - \frac{p}{\rho T}\dot \rho - \frac{\rho\mu}{T}\dot c -\frac{Z^i}{T}(\partial_i c \dot ) = \nnl{cCH2}
-&\partial_i \left[\frac{q^i - \mu j^i - Z^i\partial_kj^k/\rho}{T} + \frac{j^i}{\rho}\partial_k\left(\frac{Z^k}{T}\right)\right] + \nnl{cCH3}
q^i \partial_i\frac{1}{T} &- 
j^i \partial_i\left[\frac{\mu}{T} -  \frac{1}{\rho}\partial_k\left(\frac{Z^k}{T}\right)\right] -
\frac{\partial_i v_j}{T}\left[P^{ij} - p\delta^{ij} - Z^i \partial^j c\right].
}
\begin{table}[!h]
\caption{Entropic representation of the thermodynamic fluxes and forces of Fourier--Navier-Stokes--Cahn-Hilliard--Korteweg fluids.}
\label{sCH}
\centering
\begin{tabular}{c|c|c|c}
       &Thermal  & Diffusive &  Mechanical \\ \hline
Fluxes & $q^i$ & 
		 $j^i$ & 
	 	 $P^{ij} - p\delta^{ij} - Z^i \partial^j c$ \\ \hline
Forces & $\partial_i \left(\frac{1}{T}\right)$ & 
		 $-\partial_i\left(\frac{\mu}{T} -  \frac{1}{\rho}\partial_i\left(\frac{Z^i}{T}\right)\right)$ & 
	 	 $-\frac{\partial_iv_j}{T}$\\ 
\end{tabular}
\vspace*{-4pt}
\end{table}
 {In the right hand side of eq.~\re{cCH3}, the first term is identified as the divergence of the entropy flux and the second one is the entropy production.} One can identify the thermodynamic fluxes and forces according to Table \ref{sCH}. That can be considered as an entropic representation, because the gradients of the entropic intensives $(\frac{1}{T},-\frac{\mu}{T},-\frac{Z^i}{T})$ appear everywhere in the expression. It is notable that the flux of thermal interaction is the  {heat flux}, in spite of the modified entropy flux expression. 
The linear relation  {between} the fluxes and forces in isotropic materials leads to the following constitutive equations:
\eqn{och1}{
q^i =& \lambda \partial_i \left(\frac{1}{T}\right)  - \chi \partial_i\left(\frac{\mu}{T} -  \frac{1}{\rho}\partial_i\left(\frac{Z^i}{T}\right)\right), \nnl{oCH2}
j^i = & \chi \partial_i \left(\frac{1}{T}\right) - \zeta \partial_i\left(\frac{\mu}{T} -  \frac{1}{\rho}\partial_i\left(\frac{Z^i}{T}\right)\right)\nnl{oCH3}
P^{ij} =& p\delta^{ij}+ Z^i\partial^j c -2\eta \partial^{\langle i}v^{j\rangle} - \eta_b \partial_kv^k \delta^{ij}.
}
 {Here, the conductivity coefficients $\lambda, \chi$ and  $\zeta$ in the entropy representation are the thermal conductivity, the cross-coupling coefficient, and the thermodynamic diffusion coefficient, respectively}. The  {cross-coupling} coefficients are equal, according to Onsager reciprocity. In  {\re{oCH3}}, $\partial^{\langle i}v^{j\rangle}$ denotes the symmetric  {(deviatoric)} part of the velocity gradient, $\eta$ and $\eta_b$ are the shear and bulk viscosities, respectively.  {Let us remark that in isotropic materials the pressure tensor is symmetric since the entropy depends only on the length of the concentration gradient vector, that is on $\sqrt{\partial_ic \partial^ic}$, therefore $Z^i$ is proportional to $\partial_ic$}. The nonnegativity of the entropy production requires the following inequalities:
\eqn{sinentr}{
\lambda>0,\quad \chi>0,\quad \zeta>0,\quad \lambda\zeta-\chi^2\geq 0,\quad \eta>0,\quad \eta_b>0.
} 
It is remarkable, that for single temperature mixtures a different, reasonable flux-force representation is obtained writing the entropy balance to the following, equivalent form:
\eqn{sCH1}{
\rho \dot s + \partial_i \left[\frac{q^i - \mu j^i - (Z^i\partial_kj^k - j^i\partial_kZ^k)/\rho}{T}\right] &=\nnl{sCH2} 
	\left(q^i - \mu j^i - \frac{Z^i\partial_kj^k - j^i\partial_kZ^k}{\rho}\right) \partial_i\frac{1}{T} - 
	   &\frac{j^i}{T} \partial_i\left[\mu -  \frac{\partial_kZ^k}{\rho}\right] -
\frac{\partial_i v_j}{T}\left[P^{ij} - p\delta^{ij} - Z^i \partial^j c\right]\geq 0.
}
 {Based on the last line} of the above expression one can identify the thermodynamic fluxes and forces again. This choice is called \emph{energy representation}. It is  {notable} that the entropy flux equals the thermal flux over the temperature. Recognising that the entropy production is proportional to the reciprocal temperature, a system of fluxes and forces are given in Table \ref{tCH}. These are identical  {with the one that from} \cite{HeiEta12a}.
\begin{table}[!h]
\caption{Thermal representation of the thermodynamic fluxes and forces of Fourier--Navier-Stokes--Cahn-Hilliard--Korteweg fluids.}
\label{tCH}
\centering
\begin{tabular}{c|c|c|c}
       &Thermal  & Diffusive &  Mechanical \\ \hline
Fluxes & $q^i - \mu j^i - \frac{Z^i\partial_kj^k - j^i\partial_kZ^k}{\rho}$ & 
		 $j^i$ & 
	 	 $P^{ij} - p\delta^{ij} - Z^i \partial^j c$ \\ \hline
Forces & $-\frac{\partial_i T}{T}$ & 
		 $\partial_i\left(\mu -  \frac{\partial_kZ^k}{\rho}\right)$ & 
	 	 $-\partial_iv_j$\\ 
\end{tabular}\\
\vspace*{-14pt}
\end{table}
In isotropic materials the linear relation between the fluxes and forces results in the following constitutive equations for $q^i,j^i$ and $P^{ij}$, respectively,
\eqn{toch1}{
q^i - \mu j^i - \frac{Z^i\partial_kj^k - j^i\partial_kZ^k}{\rho} =& -\hat\lambda  \frac{\partial_iT}{T}  - \hat\chi \partial_i\left(\mu -  \frac{\partial_kZ^k}{\rho}\right), \nnl{toCH2}
j^i = & \hat\chi \frac{\partial_iT}{T}  - \hat\zeta \partial_i\left(\mu -  \frac{\partial_kZ^k}{\rho}\right),\nnl{toCH3}
P^{ij} =& p\delta^{ij}+ Z^i\partial^j c -2\eta \partial^{\langle i}v^{j\rangle} - \eta_b \partial_kv^k \delta^{ij}.
}
where the  {coefficients $\hat\lambda, \hat\chi$ and  $\hat\zeta$ are seemingly different than the previous ones. 

\subsection{Equivalence of representations} It is frequently argued, that the various flux-force representations are not equivalent \cite{Tru84b}. In our case the viscosities are the same in the two representations, but the fluxes and forces, and also the coefficients of thermodiffusion look like rather different. However, the  {\re{cCH3} and \re{sCH2}} entropy balances are the same, and with the general quasilinear solution of the inequality,-- as in our previous example,-- the coefficients are state dependent. Then, if the entropy flux does not change, a straightforward calculation gives a unique relation between the representations. For example in case of simple diffusion, without the Cahn-Hilliard extension, $Z^i\equiv 0$, and  one obtains:
\eqn{equr}{
\hat\lambda  =  \lambda + \zeta \mu^2, \qquad
\hat{\chi} = \chi -\zeta \mu, \qquad
\hat \zeta = \zeta.
}
Therefore the representations are equivalent, and the reciprocity is preserved, too. Moreover, if the inequalities in \re{sinentr} of the entropic representation are fulfilled, then similar inequalities are valid in the energy representation, too.  However, the coefficient inequalities that guarantee the nonnegative entropy production in \re{sCH2} are not sufficient to get \re{sinentr}, in partucular the firs inequality.

\subsection{Cahn-Hilliard equation} 
 {The Cahn-Hilliard equation is obtained if the density, the internal energy and the velocity fields are homogeneous and there is a linear relation between the diffusion flux and force}:
\eqn{DH}{
j^i = -\zeta\partial^i\left[\frac{\mu}{T} -  \frac{1}{\rho}\partial_k\left(\frac{Z^k}{T}\right)\right] = 
-\zeta\partial^i\left(\partial_c s - \frac{1}{\rho}\partial_k \partial_{\partial_k c}(\rho  s)\right].
}
 {It is visible} that the right hand side of the equation is proportional to the gradient of the functional derivative of the specific entropy, only if the density is constant. Then, substituting this equation into the concentration balance, one obtains the Cahn-Hilliard equation. This is a simple explanation of the  mentioned differences between the variational derivation, the Lowengrub-Truskinovsky modell \cite{LowTru98a} and the thermodynamic methods, analysed carefully in \cite{HeiEta12a}. 

 {For phase transition models, the concavity of the entropy and} the proper convexity relations for free energy are important requirements. This property ensures the stability of equilibria and the basin of attraction is related to simply connected concave regions. With additional considerations for boundary conditions the total entropy is a good candidate  {to being} a Ljapunov functional of equilibria as it is indicated in \cite{PenFif90a}.

\section{Example III: Inertial gravitation}

Recently, it was shown that a single scalar internal variable, when additively separated from the internal energy leads to a dissipative extension of classical gravitation \cite{VanAbe19m}. Now, in our third example, we extend that calculation and introduce dual internal variables in order to explore a theory of gravitation with inertia. 

We consider a single component fluid, therefore the balances of mass and energy, \re{mbal} and \re{ebal}, are the constraints for the entropy balance. Two scalar fields are introduced,  $\varphi$ and $\phi$. Our basic assumption is that the internal energy $u$ is the difference of the total energy and the energies of the gravitating matter and field:
\eqn{generg}{
u=e - \varphi - \frac{\partial_i\varphi\partial^i\varphi}{8\pi G\rho} - \frac{\psi^2}{2 K}. 
} 
Here, $G$ is the gravitational constant and $\varphi$ is the gravitational potential. It is subtracted from the specific internal energy and the square gradient of $\varphi$ represents the energy of the gravitation field \cite{Syn72a}. Additive separation of the internal variables from the energy leads to energy interpretation. The last quadratic term is responsible for inertial effects, $K$ is an inertial coefficient, as usual in variational principles. Then the specific entropy is the function also of $u$ and the specific volume $v=1/\rho$. The Gibbs relation is 
\eqn{gravGrel}{
\d e = T\d s + \left(\frac{p}{\rho^2}-\frac{\partial_i\varphi\partial^i\varphi}{8\pi G\rho^2}\right)\d\rho + \d \varphi + \left( \frac{\partial_i\varphi}{8\pi G\rho}\right)\d\partial^i\varphi +\frac{\psi}{K}\d\psi.
}
 {Similarly as before, the calculation of the entropy balance is}
 \eqn{entrbaltot}{
\rho\dot s &+\partial_i\left[\frac{1}{T}\left(q^i+\frac{1}{4\pi G}\dot\varphi\partial^i\varphi\right)\right] = \nnl{k1} &\left(q^i+\frac{\dot{\varphi}}{4\pi G} \partial^i\varphi\right)\partial_i\left(\frac{1}{T}\right) + 
\frac{\dot\varphi}{4\pi G T}\left(\partial_k\partial^k\varphi - 4\pi G\rho\right) - \frac{\rho}{KT} \psi \dot \psi\nnl{k2}
&\qquad\left[P^{ij} - p\delta^{ij} - \frac{1}{4\pi G}\left(\partial^i\varphi\partial^j\varphi -
	\frac{1}{2}\partial_k\varphi\partial^k\varphi\delta^{ij} \right) \right]\frac{\partial_iv_j}{T} \geq 0
}
The corresponding fluxes and forces are shown in Table \ref{ff}. Both the mechanical and thermal thermodynamic forces have changed due to the presence of gravitation. There is a contribution to the pressure, $P^{ij}_{grav}= \frac{1}{4\pi G}\left(\partial^i\varphi\partial^j\varphi -\frac{1}{2}\partial_k\varphi\partial^k\varphi\delta^{ij} \right)$, too. The first term, the thermal interaction is a product of vectors, the last, term, the mechanical one is a product of second order tensors. The scalar second and the third terms determine the evolution equations for the gravitational potential and for the second scalar field, $\phi$ and $\psi$, respectively. Therefore, in isotropic materials, cross-effects are possible between the scalar part of the pressure and the scalar fields. Hence the linear constitutive equations are  {
\eqn{constrel}{
q^i+\frac{\dot{\varphi}}{4\pi G} \partial^i\varphi &=-\lambda \frac{\partial_i T}{T} ,  \nl{cr_heat}
\dot\varphi &= l_1 \left(\frac{\partial_i\partial^i\varphi}{4\pi G } -\rho\right) - 
		l_{12} \rho \frac{\psi}{K} -
		l_{13} \partial_kv^k, \nl{cr_grav}
\dot\psi &= l_{21} \left(\frac{\partial_i\partial^i\varphi}{4\pi G } -\rho\right) - 	
		l_{2} \rho \frac{\psi}{K} -
		l_{23} \partial_kv^k, \nl{cr_iner}
P^i_i -3 p +\frac{\partial_i\varphi\partial^i\varphi}{8\pi G } &= 
	l_{31} \left(\frac{\partial_i\partial^i\varphi}{4\pi G } -\rho\right) - 	
		l_{32} \rho \frac{\psi}{K} -
		l_{3} \partial_kv^k, \nl{cr_bpress}
P^{ij}\! -\! P^k_k\frac{\delta^{ij}}{3}\! &-\! \frac{1}{4\pi G}\left(\partial^i\varphi\partial^j\varphi \!-\! \frac{1}{3}\partial_k\varphi\partial^k\varphi\delta^{ij}\right) 	
=	-\eta\left(\partial^iv^j\! + \partial^jv^i\! -\! \frac{2}{3}\partial_kv^k\delta^{ij}\right)\!. 
}
}
\begin{table}[!h]
\caption{Thermodynamic fluxes and forces of self-gravitating fluids with gravitational inertia. $P^{ij}_{grav}$ is the gravitational pressure.}
\label{ff}
\centering
\begin{tabular}{c|c|c|c|c}
       &Thermal  & Gravitational &  Mechanical & Gravoinertial \\ \hline
Fluxes & $q^i+\frac{\dot{\varphi}}{4\pi G} \partial^i\varphi$ & 
		 $\dot\varphi$ & 
	 	 $P^{ij} - p\delta^{ij} - P^{ij}_{grav}$ &
		 $\dot{\psi}$\\ \hline
Forces & $\partial_i \left(\frac{1}{T}\right)$ & 
		 $\frac{1}{T}\left(\frac{\partial_k\partial^k\varphi }{4\pi G}- \rho\right)$ & 
	 	 $-\frac{\partial_iv_j}{T}$&
		 $-K\rho {\psi}$\\  
\end{tabular}\\
\vspace*{-4pt}
\end{table}

Here, $l_3/3 = \eta_b$ is the bulk viscosity and $\eta$ is the shear viscosity. The classical Newtonian gravitation is obtained if there is no gravitation related dissipation, that is the thermodynamic force of gravitation is zero, that is the Poisson equation is valid and $l_{12}$, $l_{13}$, $l_{2}$ and $l_{2}$ are zero. In this case, the gravitational pressure has the following remarkable property: $\partial_jP^{ij}_{grav} = \rho \partial^i\varphi$.
Let us  {observe} the contribution of the dissipative terms in \re{cr_grav} and \re{cr_iner}, but without the coupling with divergence of velocity, that is 
\eqn{rconstrel1}{
\dot\varphi &= l_1 \left(\frac{\partial_i\partial^i\varphi}{4\pi G } -\rho\right) - 
		l_{12} \rho\frac{\psi}{K}, \nl{rconstrel2}
\dot\psi &= l_{21} \left(\frac{\partial_i\partial^i\varphi}{4\pi G } -\rho\right) - 	
		l_{2} \rho \frac{\psi}{K},
}
where $l_1>0, l_2>0$ and $l_1l_2- (l_{12}+l_{21})^2/4>0$, according to the second law. Eliminating the second field, $\psi$, one obtains the following differential equation:
\eqn{dgravde}{
K\rho\ddot \varphi - K \rho l_1\frac{d}{dt} \left(\frac{\partial_i^i\varphi}{4\pi G}- \rho\right) - 
	L \left(\frac{\partial_i^i\varphi}{4\pi G}- \rho\right)+l_2\dot \varphi = \nnl{dgravde1}
K\rho l_1\frac{d}{dt} \left(\frac{\partial_i^i\varphi}{4\pi G}-\frac{\dot\rho}{l_1} -\rho\right) - 
	LK \left(\frac{\partial_i^i\varphi}{4\pi G}-\frac{l_2}{L}\dot\varphi-\rho\right) = 0.
}
This is the governing equation  {for the} dissipative, massive Newtonian gravitation. The ideal, nondissipative case is obtained if the transport matrix is antisymmetric, i.e., $l_{1}=l_{2}=0$, and $l_{12}=-l_{21} = a$. In this case \re{dgravde1} reduces to the wave equation, as it is expected:
\eqn{dgravdewave}{
\frac{K4\pi G}{ a^2 }\rho\ddot \varphi - \partial_i^i\varphi- 4\pi G \rho = 0.
}

\section{Summary and conclusions}
\enlargethispage{20pt}
In this paper, we have surveyed the capabilities of modern thermodynamics to construct evolution equations reflecting memory and nonlocal phenomena. The focus was on the evolution equations of ideal processes, that are typically constructed from variational principles. The demonstrative examples of the previous sections considered pure memory, pure nonlocal and a mixed memory-nonlocal situations. In every case the principle behind the procedures was solely the second law and the evolution equations for ideal processes emerged in case of zero entropy production.

In Section 2 we have shown Hamiltonian dynamics of a point mass can be reproduced in a thermodynamic framework: instead of starting from a Lagrangian one can get the symplectic form of Hamiltonian mechanics and also the dissipative extension, both with direct thermodynamic arguments. Some dissipative effects are unexpected and surprising from a pure mechanical point of view. 

In Section 3 the example of a two component Fourier-Navier-Stokes-Cahn-Hilliard--Korteweg fluid demonstrated thermodynamic methods in case of a weakly nonlocal continuum. The simplest classical method of divergence separation was applied, with a straightforward generalisation of the Gibbs relation for deriving the Cahn-Hilliard equation. This equation is typically constructed with a combination of variational and thermodynamic principles. Here the balances of mass and energy are constraints for the entropy inequality. The traditional method of nonequilibrium thermodynamics reproduced the same set of equations that have corrected the variational approach of Lowengrub and Truskinovsky \cite{LowTru98a} and were obtained by rigorous arguments of rational thermodynamics \cite{HeiEta12a}. 

In Section 4 we have shown a combined  {utilization} of dual variables and divergence separation to get both memory and nonlocal equations by constructing a dissipative theory of Newtonian gravitation with inertial effects. It is remarkable, that the wave equation of the gravitational potential was only a part of the nondissipative limit of the theory: the scalar evolution equation of the gravitational potential can be coupled to the scalar part of the mechanical interaction, to the spherical part of the stress field.

These examples demonstrate, that nonequilibrium thermodynamics can substitute variational principles and construct  both nondissipative and dissipative parts of the evolution equations without any further ado. Hamilton principle emerges for the nondissipative part. For phase fields, where only the spatial part of a variational  principle is necessary the substitution is complete. 

There are some important particular aspects, where Hamilton principle from mechanics and second law of nonequilibrium thermodynamics are conceptually comparable. 
\begin{itemize}
\item Thermodynamics, is connected to the material. Regarding mechanical processes it means that the separation of material and observer motion, the concept of the material body, the comoving observer is unavoidable. That automatically requires the separation of bulk and current densities of extensive quantities. Entropy and Lagrangian both assumed to be extensive, where first the density is defined and then the flux is constructed: it is only  {implicitly included} by natural boundary conditions in variational principles or by Liu procedure based second law exploitation while it is explicit and clear with the divergence separation method. None of these techniques are unique, complete divergences (or time derivatives, gauges) can be added. 
\item Symmetries identify the conserved quantities and balance laws by Noether theorem for variational principles. Nothing similar exist in nonequilibrium thermodynamics. There the basic fields are mostly the conserved quantities from variational principle identification and the balances are given  {apriori} as constraints. That we have seen in the last two examples is actually a general approach in thermodynamics.  In this sense the two methods seem to be supplementary. This is the approach of GENERIC \cite{Ott05b}. In principle, a direct  application of symmetry requirements to the total, spacetime integrated entropy is also possible. 
\item However, then one must face conceptual problem. What could be considered as thermodynamic state variable? In this work we used fundamental physical quantities, position, momentum, gravitational potential as thermodynamic state variables. The idea whether gravity and other theories of physics are emergent or not is exactly this question \cite{Ver11a,Ver17a}. As a research program, up to now, it is focused on general relativity and authors  switch fast to statistical approaches looking for the origin of thermodynamics. However, the motivation comes from the nonrelativistic, classical field theories  and the clarification of the assumption behind emergent gravity in \cite{Ver11a} leads finally to a separation of divergences \cite{Hos10m}. 
\end{itemize}

At the end, let us outline the Hamilton variational principle for a scalar  field $\phi$ in nonrelativistic spacetime with a first order weakly nonlocal density. This is constructed from a Lagrangian, $L(\phi,\partial_t\phi,\partial_i\phi)$, and based on the next formal procedure, the `variation' of the action functional, the integral of the Lagrangian at a compact $V\times [t_1,t_2]$ spacetime domain:
\eqn{vpr1}{
\delta S(\phi) = &\delta \int_t\int_V L(\phi,\partial_t\phi, \partial_i\phi)\;\d t \d V =  \int_t\int_V \delta L(\phi, \partial_t\phi, \partial_i\phi)\;\d t \d V  = \nnl{vrpr2}
&\int_t\int_V \left[\partial_\phi L -\partial_t(\partial_{\partial_t\phi}L) -\partial_i(\partial_{\partial_i\phi}L) \right] \delta\phi\;\d t \d V + \nnl{vpr3}
&\left.\int_V \partial_{\partial_t\phi}L \delta\phi \;\d V \right]_{t_1}^{t_2}+
\int_t\oint_{\partial V} \partial_{\partial_i\phi}L \delta\phi\; \d t \d A_i =0.  
}
Then one concludes that the following Euler-Lagrange equation is valid:
\eqn{ELe}{
\partial_\phi L -\partial_t(\partial_{\partial_t\phi}L) -\partial_i(\partial_{\partial_i\phi}L) = 0.
}

When properly formulated it is a process of differentiation of $S$ on a Banach space of functions that disappear at $\partial_V\times \{t_1,t_2\}$, at a boundary of the spacetime domain. There is an affine space behind, with natural boundary conditions fixed with the last two integrals of \re{vpr3}. The Banach space derivative is zero, because of the assumed extremum property, \re{ELe} is a necessary condition. 

{As a physical principle it is a disaster. The Lagrangian is coming out of the blue, the action is not an extremum at several physical theories, initial value problems are not included, most of the fields in physics are measures and not functions, etc. Considering only some aspects of the mathematical formulation. Moreover, in most of the classical, everyday physics, the dissipative processes are excluded, evolution equations for dissipative processes cannot be generated using this variational principle without any further ado.}

One can remedy several mentioned problems. There are more general function spaces than Banach, initial value problems can be considered, some dissipative processes can be added. As a tool, it is useful, however, as a principle it is compromised. The limited validity of the approach cannot be denied. 

Up to now, the attempts to find a principal approach for constructing evolution equations for dissipative processes were based on well established methods for ideal processes. We were looking for the extensions and modifications of the variational principle \re{vpr3} keeping elements of the formal method and the idea of the extremum. In this paper we argue, that one can start from the other end and assume an inequality instead of the equality. Interestingly and importantly several methods and conditions of the exploitation of the inequality can reproduce Euler-Lagrange equations. This observation enable us to formulate a different research program. One can postpone the decision whether the fields are emergent and what kind of microscopic and submicroscopic composition of the material can lead to an entropy function. Instead we may assume that entropy, like a Lagrangian, is the generating potential of the evolution of the fields and deal with the consequences. 

This program is already running, mostly unconsciously, in the various branches of nonequilibrium thermodynamics. Any methods that construct evolution equations with thermodynamic methods are contributing. We have already mentioned here Extended Thermodynamics, GENERIC, conservation-dissipation formalism, virtual power \cite{Ott05b,MulRug98b,JouAta92b,Yon19tvm}. It is very successful in classical, nonrelativistic situations where dissipation is apparent. Then, the classical concept local equilibrium must be well specified. In our case the explicit use of Gibbs relation is that defines the concept of local equilibrium in the treated examples. The most important particular aspect is that a field theory must start from a local approach, from a density or from a specific quantity as thermodynamic potential, otherwise the separation of bulk and surface contributions, time and spatial nonlocality, that is memory and gradient effects can be problematic.

From a conceptual point of view this program is a big step, with several consequences. First of all it is a unification. Our evolution generator extremum principle, the selection rule for the laws of nature, is not for independent any more, it is of thermodynamic origin. Thermodynamic, but not statistical, the axioms are related to the stability structure, the existence of a concave and increasing potential. Whether it is the second law that is elevated to a first principle level, or the first principles are better understood as emergent? We prefer the first interpretation: the rule is to select the stable materials with the entropy as a Lyapunov functional of the equilibrium. Stability is clearly a selection rule, unstable materials survive only under special conditions. The second interpretation denies the existence of first principles. If everything is emergent, then ideal processes are exceptional, when the ever existing microlevel is somehow insulated. The fact, that the two interpretations are practically the same is a disturbing consequence. It may be reasonable to postpone the decision until more is known about the performance of this possible unification, e.g. until electrodynamics is derived from nonequilibrium thermodynamics.

\section{Acknowledgement}   
The authors thank Tamás Fülöp for valuable discussions.

The work was supported by the grants National Research, Development and Innovation Office - NKFIH 116197(116375), 124366(124508), 123815, KH130378, TUDFO/51757/2019-ITM (Thematic Excellence Program) and FIEK-16-1-2016-0007. The research reported in this paper was supported by the Higher Education Excellence Program of the Ministry of Human Capacities in the frame of Nanotechnology research area of Budapest University of Technology and Economics (BME FIKP-NANO).


\end{document}